\documentstyle[twocolumn,epsf,prl,aps]{revtex}
\begin{document}

\title{Air entrainment through free-surface cusps}

\author{
Jens Eggers
}

\address{
Universit\"{a}t Gesamthochschule Essen, Fachbereich Physik,
45117 Essen, Germany  }

\maketitle
\begin{abstract}
In many industrial processes, such as pouring a liquid or coating 
a rotating cylinder, air bubbles are entrapped inside the liquid.
We propose a novel mechanism for this phenomenon, based on the 
instability of cusp singularities that generically form on free 
surfaces. The air being drawn into the narrow space inside the cusp 
destroys its stationary shape when the walls of the cusp come too close. 
Instead, a sheet emanates from the cusp's tip, through which 
air is entrained. Our analytical theory of this instability 
is confirmed by experimental observation and quantitative 
comparison with numerical simulations of the flow equations.
\end{abstract}


\begin{figure}
\begin{center}
\leavevmode
\epsfsize=0.5 \textwidth
\epsffile{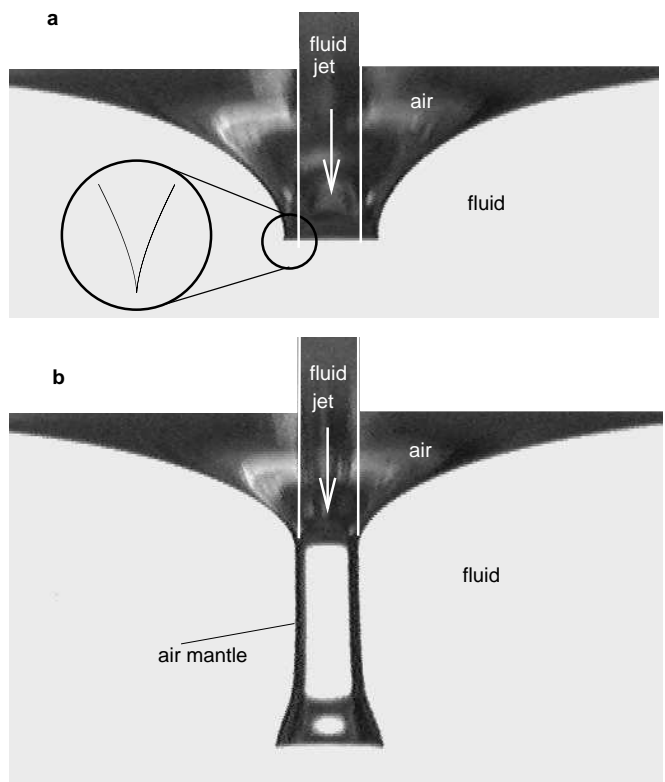}
\end{center}
\caption{
{\bf a} Cross section of the stationary air-fluid interface 
produced by a thin (1mm) stream of viscous oil poured into a 
deep pool of the same fluid. The position of the cusp is 
marked by a circle. {\bf b} A hollow cylinder of air forms after 
the cusp has become unstable at a slightly higher flow rate. (Photograph
by Itai Cohen)
}
\label{pour}
\end{figure}

Air bubbles are a ubiquitous presence in fluid flow, appearing 
when pouring a liquid into a beaker, when beating an egg, or in 
river streams. This aeration is often desirable, for example to
promote chemical reactions \cite{B93}, yet in many industrial 
processes entrainment of air bubbles is detrimental to the quality 
of the product, rendering the flow unsteady. For example, it is 
the single most important factor limiting the speed at which paints 
or coatings can be applied to a solid surface \cite{SK00,BM80}.
But in spite of its importance, no general understanding of 
air entrainment exists, except
for the rather special circumstance that the free surface conspires 
to enclose an air bubble from all sides, as was recently found for 
a disturbed water jet \cite{OOP00}. Instead, here we argue that the 
presence of cusp singularities on the free surface results in a generic 
mechanism for air entrainment. 

In recent years it has emerged that free surfaces are
rather susceptible to the formation of sharp tips 
or cusps \cite{BM80,R68,JNRR91,JM92}. This is true in particular
for viscous fluids, where shear stresses 
are strong compared with surface tension forces \cite{JM92}.
Examples are drop impact on a surface \cite{WY99}, 
jets impinging on a pool of liquid \cite{B93}, and the 
coating of a pre-wetted solid cylinder \cite{BM80}. 
Once the air enters these narrow passages, the gas flow 
serves to destroy the original structure to let the air penetrate 
below the surface. What is surprising about this novel mechanism 
is that the forces the gas flow exerts on the fluid plays the 
crucial role, even though the viscosity of the air may be smaller 
by many orders of magnitude than that of the fluid. 
Other instabilities may occur at a three-phase boundary,
for example when the solid to be coated is dry, a problem 
studied in \cite{SK00,VIC00}. In the presence of surfactant, still 
another mechanism for the instability of a cusp has been suggested 
by Siegel \cite{S99}.

For purposes of illustration, consider the particular example of
a two-dimensional cusp that forms when a thin stream of a viscous 
silicone oil is poured into a container of the same fluid. Since
the falling liquid drags other fluid away from the surface, a dip
is produced around the fluid stream. Increasing the flow rate above
a critical value, this dip is no longer smooth, but a singular point
on the surface is approached with two vertical tangents. A cross 
section of this cusped profile is shown as a black silhouette 
in Fig.\ref{pour}a, the outer wall of the free surface ending in a
vertical tangent at the cusp point. For clarity, the lighting is 
chosen such that the free surface appears opaque, so the falling jet 
is only indicated symbolically to guide the eye, but not visible 
directly. 

Increasing the flow rate still further, there is a second critical 
value where the stationary profile of Fig.\ref{pour}a ceases to 
exist and a sheet of air shoots out from the tip of the cusp. The 
bottom picture shows this dynamical structure 1/60th of a second after 
the cusp has become unstable. A thin air sheet 
now forms the wall of a transparent fluid cylinder. 
The details of this dynamical structure, such as the bell-shaped
opening at its lower rim, is not the subject of this 
paper, but only the instability of the static shape
leading to it. The cylinder eventually grows to about ten 
times the length shown, and is unstable to the formation 
of bubbles at its lower end, so the liquid pool quickly
becomes contaminated by bubbles of a broad variety of
sizes. 

\begin{figure}
\begin{center}
\leavevmode
\epsfsize=0.4 \textwidth
\epsffile{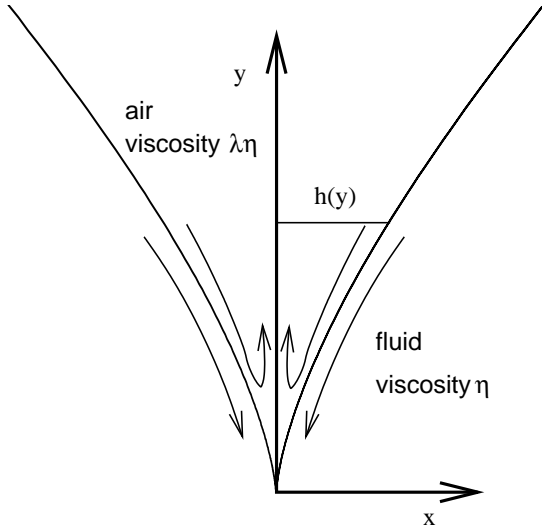}
\end{center}
\caption{
The local shape of the cusp, cut perpendicular to 
the sheet of air. The variable y is the distance 
from the tip.
}
\label{cusp}
\end{figure}
Since the air sheet near the cusp is of micron size, the 
curvature with which it is wrapped around the impinging
jet is of no consequence and the cusp can be viewed as a 
two-dimensional object. In this spirit, Joseph et al. \cite{JNRR91} 
performed experiments with a two-dimensional flow produced by
two counter-rotating cylinders submerged 
below the surface of a very viscous liquid. If the cylinders are 
placed sufficiently close to each other, a cusp forms in a 
symmetrical position between the cylinders \cite{JM92}. Letting the 
distance between the rollers going to zero, Jeong and Moffatt found
a family of exact solutions to this problem, obeying the local scaling form 
\begin{equation}
\label{local}
h(y) = \kappa^{-3/4} H(y\kappa^{1/2}),
\end{equation}
where $\kappa$ is the curvature at the tip (see Fig.\ref{cusp}). 
The structure of the solution (\ref{local}) is typical for
flows near singularities \cite{E97,SBN94,ZKFL00}, 
which involve very small scales. Physically (\ref{local}) 
means that the shape of the interface is {\it independent}
of scale, up to a rescaling of the axes.
The other crucial property of singularities is that their shape
is universal, i.e. independent of the particular type of 
flow that generates the cusp. Thus our theory, based on the 
stability of such a singular structure, will be equally general.
Indeed, Antonovskii \cite{A96} 
discovered yet another class of exact solutions, but where the cusp is
formed on the surface of a circular bubble.
A local analysis reveals that the scaling function 
\[
H(\xi) = \sqrt{a\xi}(\xi+\sqrt{2/a})
\]
is identical to the one in \cite{JM92} except for a different 
value of the numerical constant $a$, confirming the expectation 
that the flow on small scales is universal, independent of the 
particular features of the driving flow. 

In all solutions, the tip curvature $\kappa$ grows 
{\it exponentially} with the capillary number $Ca=\eta U/\gamma$, where 
$\eta$ is the fluid viscosity, $\gamma$ the surface tension, 
and $U$ is a typical velocity scale of the external flow. 
Thus as the strength of the driving flow increases relative to
the surface tension, the size of the tip may easily 
reach microscopic dimensions \cite{JM92} if the effect of
the air is not taken into account. Without it, stable solutions 
are thus predicted to exist 
for all capillary numbers, in disagreement with experiment.
Moffatt suggested \cite{M95} that this is because all previous 
analyses neglect the viscosity $\lambda\eta$ of the air being
drawn into the cusp by the flow $u_y^{(0)}(y)$ parallel 
to the cusp surface. The air entering a 
narrow space and having to escape again generates a so-called
lubrication
pressure $p_{lub}(y)$ inside the cusp, whose derivative with
respect to the distance $y$ from the cusp is 
\begin{equation}
\label{lub}
p_{lub}' = 3\lambda\eta u_y^{(0)}(y)/h^2(y)
\end{equation}
by Reynolds' theory \cite{R86}. Since the cusp narrows like 
$h(y)\sim y^{3/2}$, the lubrication pressure pushes the walls
apart according to $p_{lub}\sim y^{-2}$, just as it would keep separated
to narrowly spaced mechanical parts. 

\begin{figure}
\begin{center}
\leavevmode
\epsfsize=0.4 \textwidth
\epsffile{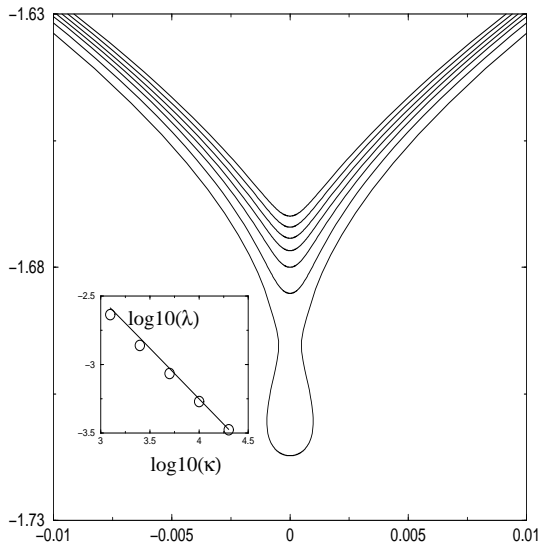}
\end{center}
\caption{
A boundary integral simulation of a bubble in the flow 
proposed by Antonovskii for $\epsilon=5$ and $Ca = 0.0992$ 
\protect\cite{A96}.
The undisturbed bubble radius is used to non-dimensionalize 
all lengths in the problem. As $\lambda$ is increased, the tip
is pushed forward, but becomes narrower at a given $y$.
The lowest profile is non-stationary.
The inset shows the critical value of $\lambda$ beyond which
there is no more stationary solution for a given curvature.
     }
\label{sim}
\end{figure}

Figure \ref {sim} proves by direct numerical simulation that 
this is enough to destroy the stationary solution found for 
$\lambda=0$. We use a boundary integral code \cite{RA78,P98}, optimized to 
resolve the cusp between two merging cylinders \cite{ELS}, neglecting
the fluid inertia. Starting from Antonovskii's solution with 
$\kappa_0=10^4$, $\lambda$ is increased in steps of $2.5\cdot10^{-5}$,
pushing the interface forward,but only every fourth  profile 
is shown. At $\lambda = 5.5\cdot10^{-4}$, no more 
stationary solution is found, but instead air enters the fluid 
forming a narrow sheet, as seen in Fig. \ref{pour} and observed in
earlier experiments \cite{M2} An important 
consequence is that in a physically correct description which 
incorporates the effect of the air (or some other gas atmosphere), 
molecular dimensions are never reached, so continuum theory
remains valid throughout \cite{S98}. 

To describe the influence of the air analytically, note that the
extra transverse velocity field $u_x^{(\lambda)}(y)$ generated by 
the air pressure can simply 
be added to $u_x^{(0)}(y)$ as given in \cite{JM92}, since Stokes' 
equation is linear. Geometrically, the cusp looks like a
two-dimensional {\it crack} entering the fluid, a problem 
well studied in linear elasticity \cite{M53}. 
Borrowing Muskhelishvili's result, we can now write
$u_x^{(\lambda)}(y)$ as
\begin{eqnarray}
\label{kernel}
&& u_x^{(\lambda)}(y) = \int_0^{\infty} p(y')m(y'/y)dy',\\
&& m(x) = (1/2\pi)\ln((1+\sqrt{x})/(1-\sqrt{x})). \nonumber
\end{eqnarray}

But our free-surface problem is of course non-linear, since the
free surface has to follow the streamlines of the flow, which are
modified by $u_x^{(\lambda)}$. Namely, the inverse slope 
of the interface is 
\begin{equation}
\label{stream}
h' = (u_x^{(0)} + u_x^{(\lambda)}) / u_y^{(0)},
\end{equation}
where $u_x^{(0)}$ and $u_y^{(0)}$ are known \cite{JM92} and 
$u_x^{(\lambda)}$ is calculated from $h$ as outlined above.
Note that while $u_x^{(0)}$ has to point inward towards the 
cusp, $u_x^{(\lambda)}$ results from the lubrication pressure and
points away from the cusp. Thus, at a given distance $y$ from the tip,
$h'$ becomes smaller and the channel {\it narrows}.
Owing to (\ref{lub}) the lubrication pressure is increased,
further increasing $u_x^{(\lambda)}$, so this nonlinear feedback 
eventually destroys the cusp solution, as seen in Fig. \ref{sim}.

It is extremely useful to recast equations (\ref{lub})-(\ref{stream})
in the scaling variable $\xi=y\kappa^{1/2}$, cf. (\ref{local}).
First, from (\ref{stream}) and since $u_y^{(0)}$ is 
a constant up to logarithmic corrections,
$u_x^{(0)}$ must scale like $\kappa^{-3/4}\kappa^{1/2}=
\kappa^{-1/4}$. From (\ref{lub}) $p_{lub}$ is estimated as
$p_{lub}\sim\lambda\kappa$, and thus 
$u_x^{(\lambda)}\sim\lambda\kappa^{1/2}$ from integrating once. 
The two opposing velocities become comparable at some critical
value of the parameter $r = \lambda\kappa^{3/4}$. 
Thus (\ref{lub})-(\ref{stream}) can be recast in similarity variables,
leading to an integral equation for the correction $H_c(\xi)$
to the unperturbed surface profile $H(\xi)$:
\begin{equation}
\label{int}
H_c(\xi)=-\frac{3r\xi}{u_y^{(0)}(\xi/\kappa^{1/2})}
\int_0^{\infty}\frac{u_y^{(0)}(\eta/\kappa^{1/2})M(\eta/\xi)}
{\left[H(\eta) + H_c(\eta) \right]^2} d\eta
\end{equation}
where $M'(\eta) = m(\eta)$. It is a simple matter to solve 
(\ref{int}) numerically, giving increasingly large corrections
$H_c(\xi)$ to the profile as $r$ is raised. Since $H_c$ is negative,
the denominator in the integrand of (\ref{int}) {\it decreases}, 
leading to a further increase in the absolute magnitude of the
correction, in accordance with the qualitative argument given above. 
Owing to this nonlinear feedback, a solution ceases to exist 
above a critical value of $r$, which has a weak dependence 
on $\kappa$ due to the logarithmic dependence of $u_y^{(0)}$
on its argument. Hence for the flow parameters of Fig. \ref{sim},
we predict that the cusp becomes unstable when the curvature reaches a 
critical value of $\kappa_{cr}\approx 0.45\lambda^{-4/3}$. This 
approximation is hardly distinguishable from the result of the 
full solution of (\ref{int}), which in the insert of Fig. \ref{sim}
is seen to be in good agreement with numerical simulations for
various values of $\lambda$. Because of the 
relationship between curvature and capillary number, this translates 
into the anticipated critical value $Ca_{cr}$ above which stationary 
solutions are no longer found. At low viscosities, 
the capillary number never even reaches
the critical value for the {\it formation} of a cusp,
so an unperturbed water jet does not entrain air \cite{OOP00}.

In conclusion, we have incorporated the effect of an outer fluid 
like air into the theoretical description of a cusp. This allows
for the first quantitative description of air entrainment 
through surface singularities. A description of the resulting sheet 
of air and its stability remains to be done. Other future 
challenges include the analysis of its close three-dimensional 
relatives, namely jet formation out of bubble tips ``tip-streaming''
\cite{S84}, Taylor cones ``electric jets'' \cite{B88}, or spouts
formed by planar interfaces ``selective withdrawal'' \cite{SG93}.

\acknowledgements
I am very grateful to Keith Moffatt for pointing
out this problem to me, and to Itai Cohen for 
donating his experimental pictures. Thanks are also due to 
Todd Dupont for help with the numerics, and to Howard Stone
for very useful discussions.

\end{document}